\documentclass[fleqn,twoside]{article}%
\usepackage{amsmath}
\usepackage{amsfonts}
\usepackage{amssymb}
\usepackage{graphicx}
\usepackage{cite}
\def\be{\begin{equation}}

\def\ee{\end{equation}}

\def\bi{\bibitem}

\begin{document}

\title{Stability of a scalar potential under gravity.}

\author{Subenoy Chakraborty $^1$ and Abhik Kumar Sanyal$^2$}
\maketitle

\noindent

\begin{center}
$^1$ Dept of Mathematics, Jadavpur University, India - 700032\\
$^2$ Dept. of Physics, Jangipur College, Murshidabad, India - 742213\\
\end{center}

\begin{abstract}
\noindent In the present note we show that the thin-wall approximation is not an assumption rather a necessary condition for satisfying Boucher's criteria for gravitational stability of the scalar potential.
\end{abstract}

\section{Introduction:}
Classical field theory admits two stable homogeneous ground states $\phi = \phi_+$ and $\phi = \phi_-$, for a single scalar field $V(\phi)$ whose action is,

\be S_1 = \int \left[{1\over 2}(\partial _\mu\phi)^2 - V(\phi)\right] d^4 x.\ee
In quantum field theory however, the ground state of higher energy $\phi = \phi_+$ corresponds to a false vacuum whose pressure is equal to the negative of energy density ($p = -\rho$). The other ground state of lower energy $\phi = \phi_-$ is called the true vacuum, as it is truly stable. Due to the energy difference between the true and the false vacua, a quantum fluctuation would cause the Higgs' field to tunnel through the energy barrier, nucleating a bubble of the broken symmetry phase. The bubble then grows at a speed approaching that of the light and the false vacuum is converted into the broken symmetry phase. This decay of the false vacuum to the true vacuum also holds in the semiclassical theory. In inflationary model \cite{1, 2, 3, 4}, the rate at which bubbles form is assumed to be very low.\\

The action for the scalar field being minimally coupled to the gravity is

\be S = \int\left[{1\over 2} g^{\mu\nu}\partial_\mu\phi\partial_\nu\phi - V(\phi) - {R\over 16\pi G}\right]\sqrt{-g} d^4~x.\ee
Coleman and Deluccia have shown \cite{5} that in the gravitational theory if the energy splitting ($\epsilon = V(\phi_+) - V(\phi_-)$) between the false and the true vacuum is small, then the tunnelling rate vanishes provided $V(\phi_+) \le 0$. In other words, gravitational effect can stabilize a scalar field configuration, which is unstable in the absence of gravity.\\

\
Now, on the basis of Witten's positive energy theorem \cite{6}, which essentially demonstrates the stability of super-symmetric solutions of super-gravity theories, Boucher \cite{7} investigated the stability of non super-symmetric solutions of super-gravity theories along with the solution of theories which do not have super-gravity extensions. The criteria to test the stability of a given configuration presented by Boucher \cite{8} is as follows. If one considers a potential $V(\phi)$ with an extremum $V'(\phi_+) = 0$, then $\phi = \phi_+$ is a stable configuration, if there exists a real function $f(\phi)$ satisfying the following boundary conditions,

\be f(\phi_+) = \left[-V(\phi_+)\over 3\kappa\right]^{1\over 2},\ee  
and  

\be 2 f'^2 - 3\kappa f^2 = V(\phi),\ee
where, $\kappa = 8\pi G$ and prime denotes derivative with respect to $\phi$.  Equation (3) implies that $V(\phi_+) \le 0$, and so using Witten's positive energy theorem \cite{6}, one can show that the state $\phi = \phi_+$ is a stable state both classically and semi-classically. \\

As already stated, Coleman and DeLuccia \cite{5} were the first to observe that when a scalar potential is coupled to gravity, tunnelling rate vanishes and the stability of the scalar potential is found under certain conditions, viz. $V(\phi_+) \le 0$. Later, Park and Abbott \cite{8} also derived the condition under which gravity can stabilize a scalar field, following Boucher's formalism \cite{7}. In both of these formalisms \cite{5, 8} thin wall approximation has been treated as an assumption. This approximation, as already mentioned is valid only in the limit of small energy density difference, viz, $\epsilon = V(\phi_+) - V(\phi_-)$ between the true and the false vacua. However, as pointed out by Park and Abbott \cite{8}, the thin wall approximation does not restrict the proof of instability when Boucher's condition \cite{7} are not met. This is because, the equality appearing in Boucher's condition \cite{7} can be replaced by an inequality, implying that any potential which is everywhere smaller (except at the false vacuum point) than an unstable potential, must itself be unstable. Since the this wall approximation is always valid in the marginal case it can be used, and then from Boucher's condition \cite{7} the instability of other potentials for which the thin wall approximation does not apply, can be inferred.\\

In the present note, we shall show that the thin wall approximation is not an assumption, but the necessary conditionrequired to satisfy Boucher's criteria \cite{7} for the gravitational stability of a scalar potential.

\section{Stability condition:}

Following Coleman and De-Luccia \cite{5}, we shall also start here by constructing a bounce. Bounce is a solution of the Euclidean field equations which satisfy certain boundary conditions and have finite Euclidean action. It is also $O(4)$ invariant. Thus,we can write the most general rotationally symmetric Euclidean metric as,

\be ds^2 = d\xi^2 + \alpha^2(\xi) d\Omega)_3^2,\ee
where, $d\Omega_3$ is the measure of solid angle on the three-sphere, and $\alpha(\xi)$ is the radius of curvature of each three-sphere, i.e., the Euclidean analogue of Robertson-Walker scale factor. Considering $\phi$ to be a function of $\xi$ only, we can express the scalar field equation as,

\be \ddot \phi + 3{\dot\alpha\over \alpha}\dot\phi - V'(\phi) = 0,\ee
and the only surviving compnent of Einstein's equation $R_{\mu\nu} - {1\over 2} g_{\mu\nu} R = - \kappa T_{\mu\nu}$ is $R_{\xi\xi} - {1\over 2} g_{\xi\xi} R = - \kappa T_{\xi\xi}$, which leads to the following equation,

\be \left(\dot\alpha\over \alpha\right) = {1\over \alpha^2} + {\kappa\over 3} \left({1\over 2}\dot\phi^2 - V(\phi)\right).\ee
In the above equations, an over-dot denotes derivative with respect to $\xi$ and prime denotes derivative with respect to $\phi$, as mentioned earlier. We now define a function $f(\phi)$ in terms of the Euclidean energy of the scalar field as,

\be  {1\over 2}\dot\phi^2 - V(\phi) = 3\kappa f^2(\phi).\ee
For a bounce solution, $\phi(\xi)$ is monotonic in $\xi$ and hence one can express the energy as a function of $\phi$ only. So differentiating equation (8) with respect to $\xi$ we obtain,

\be \dot\phi \ddot\phi - V'(\phi) \dot\phi = 6\kappa f(\phi) f'(\phi) \dot\phi,\ee
or 

\be \ddot\phi - V'(\phi) = 6\kappa f(\phi) f'(\phi),\ee
for $\dot\phi \ne 0$. In view of equation (9) equation (6) may now be expressed as,

\be {\dot\alpha\over \alpha}\dot\phi + 2\kappa f(\phi) f'(\phi) = 0.\ee
Again in view of equations (8) and (10), equation (7) takes the following form,

\be {4\kappa^2 f(\phi)^2 f'(\phi)^2\over 6\kappa f(\phi)^2 + 2 V(\phi)} = {1\over \alpha^2} + \kappa^2 f(\phi)^2,\ee
which may further be expressed as,

\be {2f'(\phi)^2\over 1 + {1\over \kappa^2 \alpha^2 f(\phi)^2}} - 3\kappa f(\phi)^2 = V(\phi).\ee
Equation (11) reduces to the Boucher's condition (4), if we consider 

\be {2f'(\phi)^2\over 1 + {1\over \kappa^2 \alpha^2 f(\phi)^2}} = 2f'(\phi)^2.\ee
Equation (12) is satisfied for two cases. Either if we consider $\alpha$ is large enough compared to the characteristic range of the variation of $\phi$, so that ${1\over \kappa^2 \alpha^2 f^2}$ term may be neglected, or if $f'(\phi) = 0$, at a point where $\phi = \phi_+$. The first case is clearly the thin-wall approximation, while the second case i.e., $f'(\phi) = 0$ at $\phi = \phi_+$ implies

\be V(\phi_+) = - 3\kappa f(\phi)^2,\ee
so that $V'(\phi_+) = 0$. Equation (13) is nothing but the other condition presented by Boucher viz. (3). However, equation (13) as mentioned, is only true at one particular point $\phi = \phi_+$, whereas Boucher's condition \cite{7} must hold for all values of the scalar field. Thus equation (11) only reduces to the Boucher's condition \cite{7} everywhere, provided the thin-wall approximation is valid. This means thin-wall approximation is a necessary condition for the stability of a scalar potential.

\section{Concluding remarks:}

We must confess that in the present note, no new result has been produced, other than the one already been obtained by Park and Abbott \cite{8}. Rather, we have just proved that the thin-wall approximation, that has so far been considered as an assumption, should now be viewed as the necessary condition required to obtain the gravitational stability of the scalar potential.

\end{document}